\documentclass[aps,prl,twocolumn,showpacs,amsmath,amssymb]{revtex4-2}

\usepackage{graphicx}
\usepackage{hyperref}
\usepackage{bm}
\usepackage{wasysym}
\usepackage{float}

\begin{document}

\title{Radiative corrections to inverse beta decay: a precision analysis for reactor neutrinos}

\author{Oleksandr Tomalak \thanks{tomalak@itp.ac.cn}}
\affiliation{Institute of Theoretical Physics, Chinese Academy of Sciences, Beijing \& 100190, People’s Republic of China}

\date{\today}

\begin{abstract}
We present a complete calculation of radiative corrections to the inverse beta decay reaction, $\overline{\nu}_e + p \rightarrow e^+ + n$, at reactor antineutrino energies using heavy-baryon chiral perturbation theory. Our analysis consistently incorporates quantum electrodynamics, chromodynamics, and electroweak contributions for the first time within this framework. We provide updated, high-precision cross-section predictions with a full error budget and present the positron energy spectrum, including radiative effects. The results are essential for normalizing the reactor antineutrino flux, determining neutrino oscillation parameters with subpercent accuracy, and searching for new physics at nuclear power plants.
\end{abstract}

\maketitle

\section{Introduction}

Inverse beta decay (IBD) is the primary detection channel for reactor antineutrinos, enabling precision measurements of neutrino oscillation parameters and searches for new physics. Its distinctive prompt-delayed MeV-photon time coincidence signature enables efficient background suppression. The IBD cross section relates the antineutrino flux to observable event rates~\cite{Huber:2011wv,Mueller:2011nm} and serves as key input for extracting neutrino oscillation parameters in medium-baseline reactor experiments. These measurements have already provided the first definitive result for the oscillation parameter $\theta_{13}$~\cite{DayaBay:2018yms} and achieved percent-level precision for $\theta_{13}$ and the squared neutrino mass difference $|\Delta m^2_{32}|$. Building on these successes, the Jiangmen Underground Neutrino Observatory (JUNO)~\cite{JUNO:2015zny,JUNO:2021vlw} aims to extract solar neutrino oscillation parameters, both the mixing angle and squared mass difference, with subpercent precision and to determine the neutrino mass hierarchy with $3\sigma$ significance. Notably, the first measurement of reactor neutrino oscillations by JUNO has improved the accuracy of the solar oscillation parameters by a factor $\sim 1.6$~\cite{JUNO:2025gmd,JUNO:2025fpc}. Achieving such high precision requires accurate incorporation of IBD cross sections. Although oscillation analysis can eliminate cross-section uncertainties in simplified models of straight-line oscillations, the reality is more complex: both JUNO and predominantly JUNO-TAO detectors observe antineutrinos mainly from two different nuclear power plants at varying distances from the reactor core. Consequently, IBD cross sections and associated uncertainties cannot be fully excluded in oscillation analyses, particularly in the JUNO spectral analysis. In this context, percent-level quantum electrodynamics (QED) radiative corrections must be consistently incorporated to avoid bias in cross sections, experimental observables, and extracted oscillation parameters such as $\theta_{12},~\theta_{13}$, $\Delta m^2_{21}$, and $\Delta m^2_{32}$. Furthermore, all percent-level effects are important for searches for new physics, description of antineutrino fluxes, and calibration of energy spectra at nuclear power plants.

Radiative corrections and various contributions to IBD cross sections have been evaluated over several decades in Refs.~\cite{Sirlin:1977sv,Dicus:1982bz,Vogel:1983hi,Fayans:1985uej,Bardin:1983yb,Bardin:1985fg,Seckel:1993dc,Towner:1998bh,Vogel:1999zy,Horowitz:2001xf,Kurylov:2001av,Kurylov:2002vj,Strumia:2003zx,Fukugita:2004cq,Fukugita:2005hs,Vogel:2006sg,Raha:2011aa,Leitner:2006ww,Hayes:2016qnu,Ankowski:2016oyj,Giunti:2016vlh,Ivanov:2017ifp,Ricciardi:2022pru,Altarawneh:2024sxo}. In Ref.~\cite{Sirlin:1977sv}, QED contributions were first included after accounting for high-energy electroweak and quantum chromodynamics (QCD) corrections with the treatment of ``inner" radiative corrections by the current algebra approach, which works with the full field and interaction contents of the Standard Model. Later on, an alternative consistent treatment of electroweak and QCD radiative corrections as short-distance contributions in a controlled low-energy effective field theory, the heavy-baryon chiral perturbation theory (HBChPT), was proposed in Refs.~\cite{Ando:2004rk,Raha:2011aa}. HBChPT provides a systematic expansion in powers of $\frac{q_{\rm ext}}{m_\pi}$ and $\frac{m_\pi}{\Lambda_\chi}$, with the characteristic momentum transfer in the process $q_{\rm ext}$, the pion mass $m_\pi$, and the chiral scale $\Lambda_\chi = 4 \pi F_\pi \sim m_N \approx 1~\mathrm{GeV}$, where $F_\pi$ is the pion decay constant and $m_N$ is the nucleon mass. Although QED radiative corrections were formulated within HBChPT in Refs.~\cite{Ando:2004rk,Raha:2011aa}, numerical evaluations were not carried out due to the lack of an appropriate determination of the low-energy coupling constants (LECs). Achieving the subpercent precision goals of JUNO, however, requires a consistent inclusion of short-distance contributions to the LECs. Furthermore, because JUNO's precision relies heavily on reconstructing the incoming antineutrino energy, percent-level control over the kinematics of the positron and photon in the final state demands an evaluation of radiative corrections beyond existing results in the literature.

In this work, we present the first consistent calculation of QED, electroweak, and QCD radiative corrections in IBD using HBChPT. We benefit from a recent systematic effective field theory (EFT) determination of the low-energy vector coupling constant $g_V$ in Ref.~\cite{Cirigliano:2023fnz}. Beyond the consistent EFT separation of QED radiative corrections that improves the accuracy of cross-section predictions, we update the evaluation of bremsstrahlung contributions by integrating over phase space fully analytically without the zero-recoil approximation in the final-state kinematics. Nevertheless, this approach violates the rigorous EFT power counting; it allows to capture the leading enhancements near kinematic endpoints. We perform the calculation by applying the technique from Ref.~\cite{Ram:1967zza}, further developed in Refs.~\cite{Tomalak:2019ibg,Tomalak:2021lif,Tomalak:2021hec,Tomalak:2022xup,Tomalak:2022uwv}. This allows us to provide the positron energy spectrum for the first time, complementing improvements over the previous works in the energy spectrum with respect to the sum of the positron and photon energies. Additionally, we provide a few two- and three-dimensional distributions to properly account for the effects of a radiated photon on detector signatures.

\section{Inverse beta decay}

We study the charged-current antineutrino-proton elastic scattering, $\overline{\nu}_e + p \rightarrow e^+ + n$, assuming protons at rest in the laboratory frame. We denote the proton velocity four-vector as $v^\rho$ and the spin operator as $S^\rho$. The typical reactor antineutrino energy $E_{\overline{\nu}_e}$ is below $E_{\overline{\nu}_e} \lesssim 10~\mathrm{MeV}$, allowing us to consider nucleons as heavy fields and work in the effective field theory without pions. All electroweak and QCD contributions from energy scales above the $1$-$10$ MeV scale of the experiment are included in the low-energy vector $g_V$ and axial-vector $g_A$ coupling constants of the effective Lagrangian $\mathcal L$~\cite{Ando:2004rk,Falkowski:2021vdg,Cirigliano:2022hob}:
\begin{equation} \label{eq:Lagrangian_at_leading_order}
	\mathcal L = - \sqrt{2} G_F V^\star_{ud} \overline{\overline{\nu}}_e \gamma_\rho \overline{e} \cdot \overline{n} \left( g_V v^\rho - 2 g_A S^\rho \right) p + \mathrm{h.c.},
\end{equation}
with the scale-independent Fermi coupling constant $G_F = 1.1663787(6)\times 10^{-5}~\mathrm{GeV}^{-2}$~\cite{Fermi:1934hr,Feynman:1958ty,vanRitbergen:1999fi,MuLan:2012sih} and the Cabibbo-Kobayashi-Maskawa (CKM) quark mixing matrix element $V_{ud}$~\cite{Cabibbo:1963yz,Kobayashi:1973fv,ParticleDataGroup:2020ssz}. $p$ and $n$ are the proton and neutron heavy-fermion fields. $\overline{e}$ and $\overline{\nu}_e$ denote the positron and antineutrino fields, respectively. The same Lagrangian describes another precisely measured low-energy charged-current reaction with nucleons: the neutron decay~\cite{Ando:2004rk,Tomalak:2023xgm,Cirigliano:2023fnz,Cirigliano:2024nfi}.

We evaluate IBD cross sections and all types of corrections starting from the Lagrangian in Eq.~(\ref{eq:Lagrangian_at_leading_order}). For numerical estimates, we use LECs at the chiral renormalization scale $\mu_\chi$ near the electron mass $m_e$. We take values from Refs.~\cite{Marciano:2005ec,Seng:2018qru,Seng:2018yzq,Czarnecki:2019mwq,Hayen:2020cxh,Shiells:2020fqp,Cirigliano:2023fnz,Moretti:2025qxt} for the vector coupling constant $g_V(\mu_\chi = m_e) = 1.02499(13)$. The ratio of axial-vector-to-vector coupling constants is $\lambda \equiv g_A / g_V = 1.2754(13)$, taken from PDG~\cite{Beck:2019xye,ParticleDataGroup:2024cfk,Beringer:2024ady,Workman:2022ynf} and extracted mainly from measurements of the correlation between the neutron spin and electron momentum in polarized neutron decay. To avoid uncertainties from discrepancies in neutron lifetime measurements~\cite{Wietfeldt:2018upi,Czarnecki:2018okw,Castelvecchi:2021hec,Gardner:2023wyl}, we use $g_A$-independent extractions from superallowed nuclear beta decays for the CKM matrix element. This gives $V_{ud} = 0.97348(31)$~\cite{Hardy:2020qwl}. Recent studies~\cite{Gorchtein:2018fxl,Seng:2018qru,Cirigliano:2024rfk,Cirigliano:2024msg} suggest that the uncertainty associated with this extraction of $V_{ud}$ could be underestimated.

\section{QED radiative corrections}

\begin{figure}[H]
	\centering
	\includegraphics[scale=0.2]{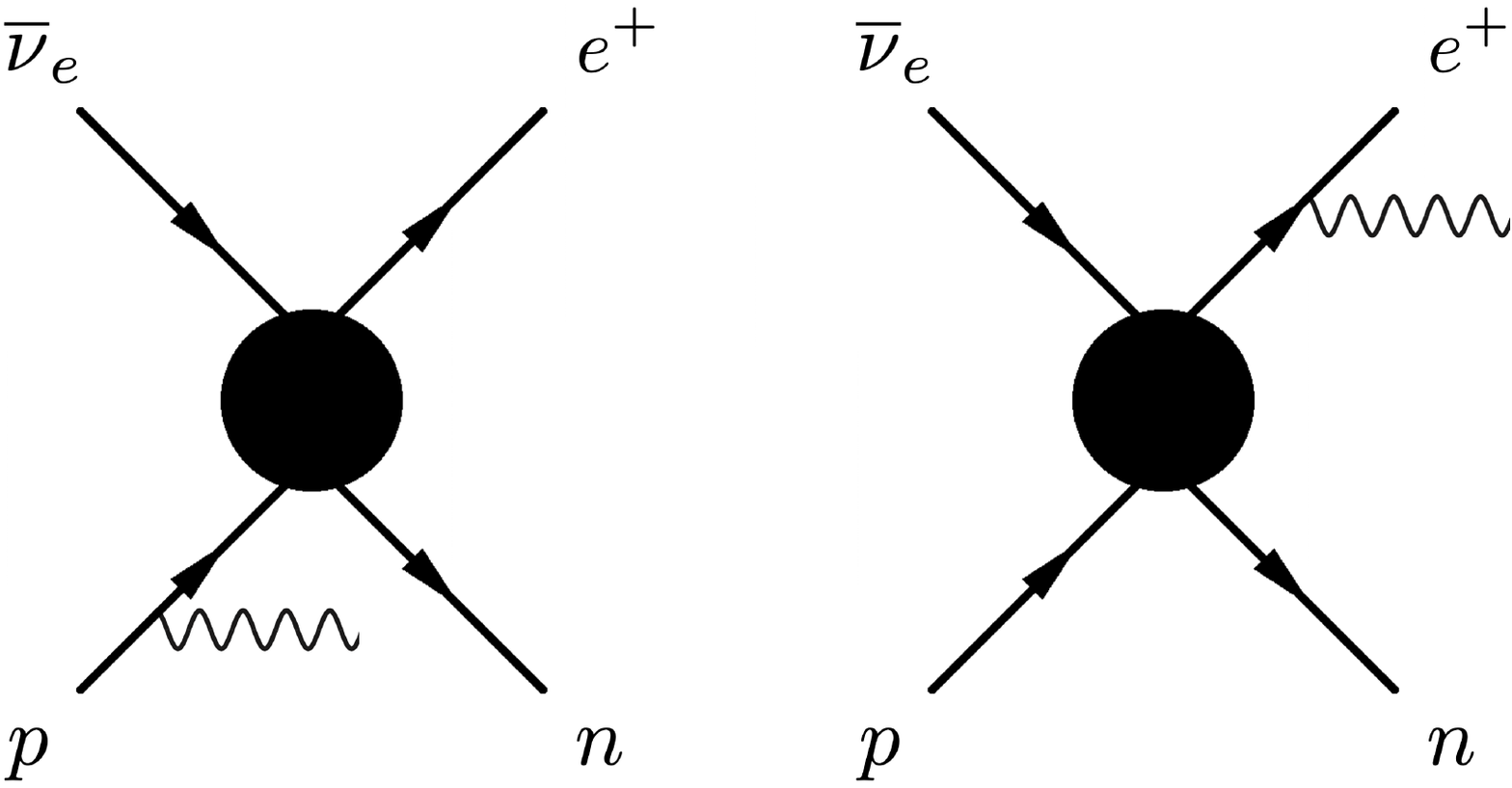}
	\caption{Leading-order bremsstrahlung diagrams for $\bar{\nu}_e + p \rightarrow e^+ + n + \gamma$. \label{fig:bremsstrahlung_graphs}}
\end{figure}

In applications with reactor antineutrinos, it is sufficient to expand the effective Lagrangian in the electromagnetic coupling constant $\alpha$ and to account for the well-known leading recoil and nucleon-structure corrections, neglecting radiative-recoil and chirally suppressed contributions. The corresponding leading in nucleon recoil QED radiative corrections can be formulated as the evaluation of the virtual photon exchange diagram in Fig.~\ref{fig:one_loop_QED} below, accounting also for the external field renormalization factors and inclusion of the radiation of one real photon by an electrically charged positron and proton; cf. Fig.~\ref{fig:bremsstrahlung_graphs}.
\begin{figure}[H]
	\centering
	\includegraphics[scale=0.11]{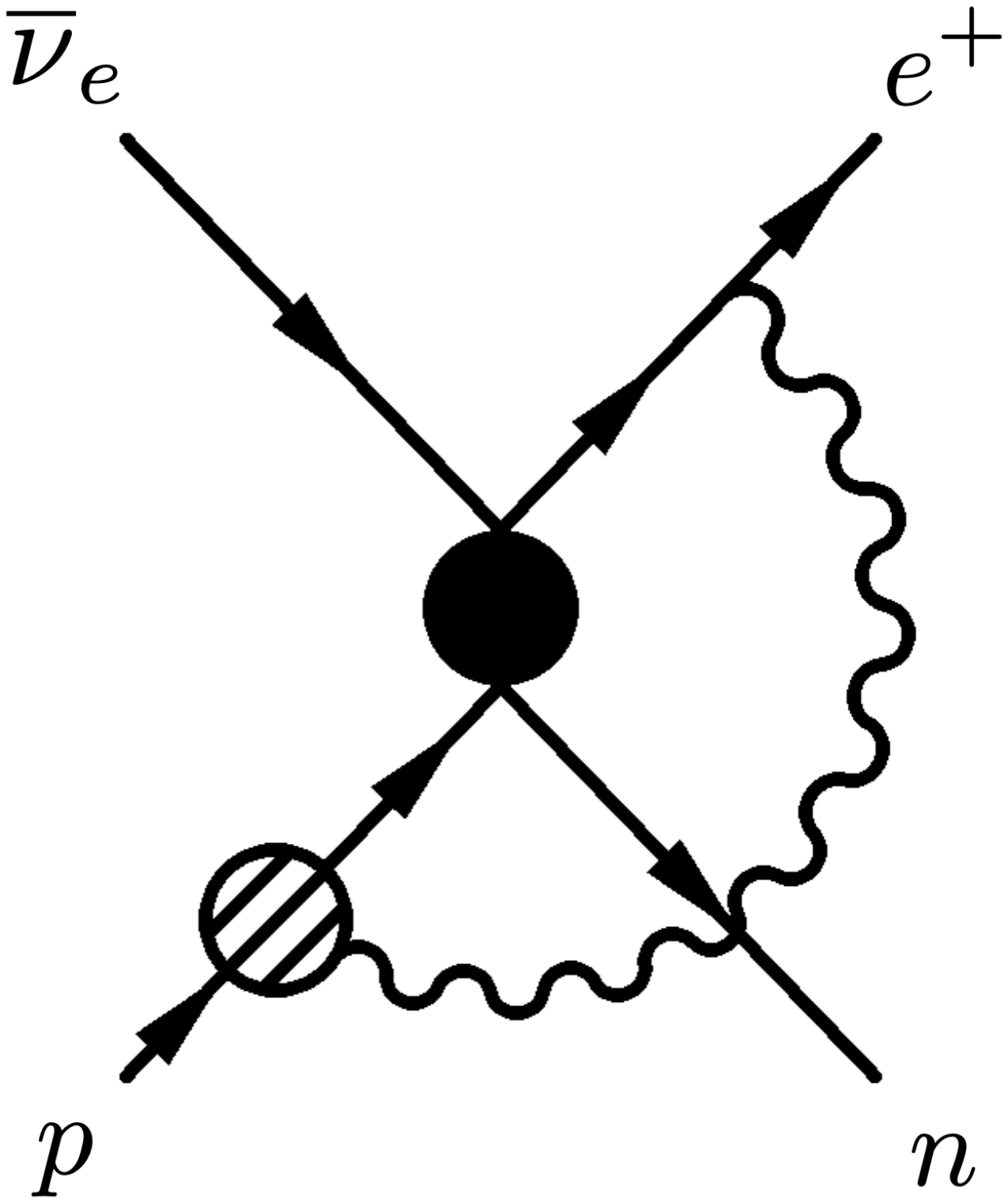}
	\caption{One-loop virtual QED correction in IBD. The photon couples to the proton and positron lines. \label{fig:one_loop_QED}}
\end{figure}
The recoil positron energy in radiation-free IBD can be determined by the incoming antineutrino energy up to $1/m_N$ corrections, which allows us to reconstruct the antineutrino energy in each interaction occurrence. The radiated photon can take away some energy from the positron and result in lower positron energies down to its rest mass. Therefore, precise measurements have to include radiative inverse beta decay events as part of the event simulation to properly account for the nonlinearity and particle dependence of energy responses~\cite{DayaBay:2019fje,JUNO:2025gmd,JUNO:2025fpc}.

We work at $\mathcal{O}(\alpha)$ in QED and include leading $1/m_N$ contributions. We perform the calculation of radiative corrections in the $\overline{\mathrm{MS}}_\chi$ renormalization scheme~\cite{Gasser:1983yg,Cirigliano:2023fnz,Cirigliano:2024nfi}. We use a photon mass regulator for infrared divergences, with cancellation between virtual and real corrections. Ultraviolet divergences are absorbed into renormalized LECs $g_V$ and $g_A$ of the effective Lagrangian in Eq.~(\ref{eq:Lagrangian_at_leading_order}).

The virtual QED corrections~\cite{Raha:2011aa} and external field renormalizations coincide with those in the neutron beta decay, besides the absence in IBD of the well-known Coulomb corrections that can be expressed in terms of the nonrelativistic Fermi function~\cite{Gamow:1928zz,Sommerfeld:1931qaf,Fermi:1934hr,Konopinski:1935zz,Morita:1963zz,Wilson:1968pwx,Halpern:1968zz,Halpern:1970it,Wilkinson:1982hu,Hoang:1997sj,Hoang:1997ui,Czarnecki:1997vz,Beneke:1999qg,Hoferichter:2009gn,Matsuzaki:2012qb,Matsuzaki:2013twa,Cirigliano:2023fnz,Hill:2023acw}. We also reproduce the known results for the leading in recoil real radiation at the level of the squared matrix elements from Refs.~\cite{Sirlin:1967zza,Garcia:1978bq,Fukugita:2004cq,Ando:2004rk,Bernard:2004cm,Raha:2011aa,Gardner:2012rp}, prior to performing the phase-space integrations. The analytic expression for the bremsstrahlung in IBD is known from applications to the neutron decay under the assumption of no recoil for the nucleon. We avoid the static-limit approximation used in previous works, perform phase-space integration in radiative IBD without kinematic approximations, and obtain slightly different results at the end-point regions. We provide all steps of the evaluation, detailed analysis, thorough comparison to prior calculations, and analytic expressions for radiative cross sections and energy spectra in our accompanying paper~\cite{Tomalak:2025jtn}.

\section{Results and discussion}

\begin{figure}[H]
	\centering
	\includegraphics[width=0.35\textwidth]{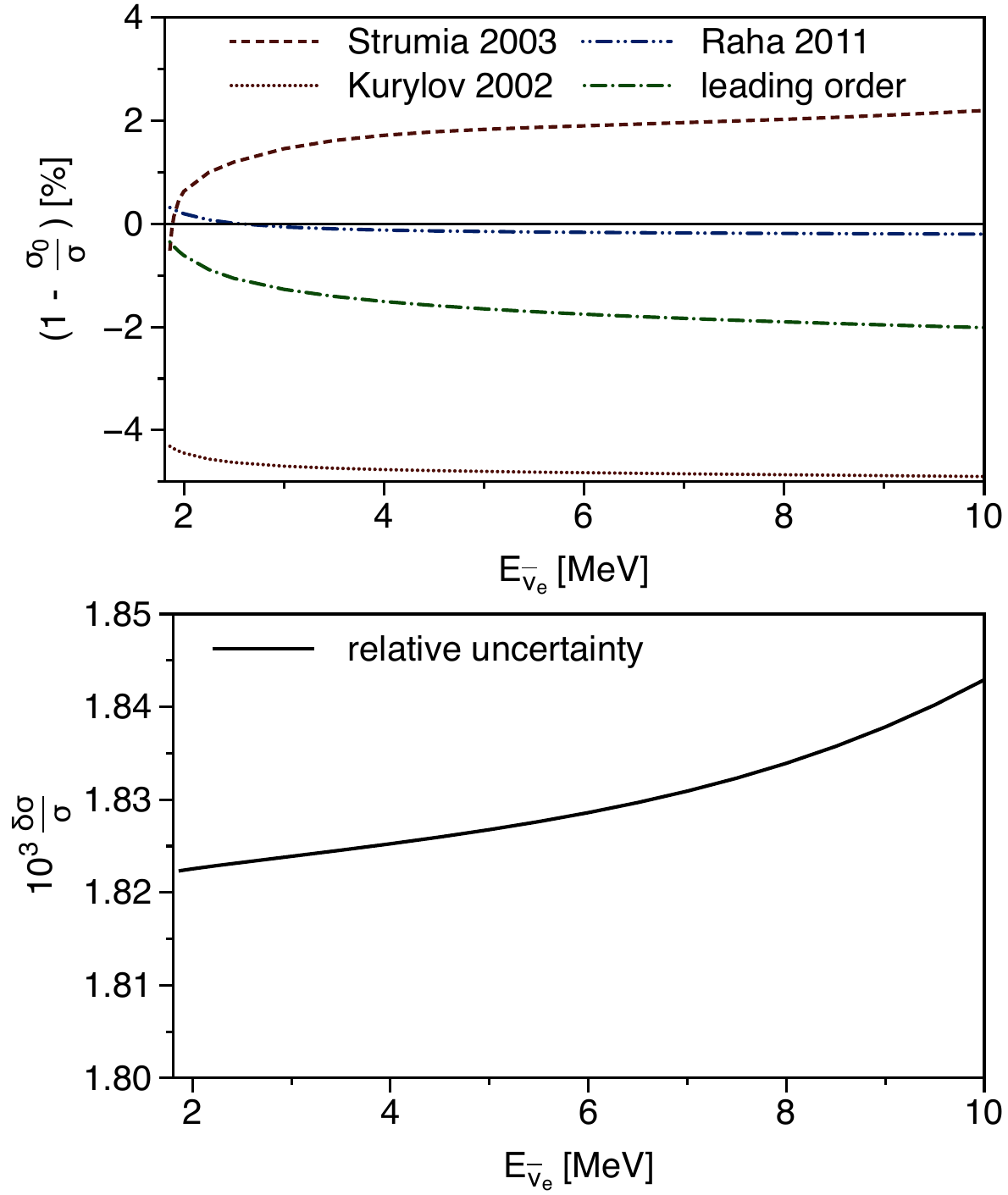}
	\caption{Upper panel: Relative difference between the total IBD cross section $\sigma$ in this work and previous calculations $\sigma_0$ is presented as a function of the antineutrino energy $E_{\overline{\nu}_e}$.  Lower panel: Relative theoretical uncertainty. Our result is compared to the cross-section parameterization in Ref.~\cite{Strumia:2003zx}, labeled as ``Strumia 2003", predictions based on the treatment of radiative corrections in Ref.~\cite{Kurylov:2002vj}, as explained in Refs.~\cite{Strumia:2003zx,Hayes:2016qnu}, labeled as ``Kurylov 2002", and radiative corrections from Refs.~\cite{Fayans:1985uej,Vogel:1999zy,Fukugita:2004cq,Raha:2011aa}, labeled as ``Raha 2011". \label{fig:total_cross_section}}
\end{figure}

First, we present the relative difference in the total IBD, $\overline{\nu}_e + p \rightarrow e^+ + n + (\gamma)$, cross section between our result $\sigma$, accounting for all effects at half-a-permille level, and other results $\sigma_0$ as a function of the antineutrino energy $E_{\overline{\nu}_e}$ in Fig.~\ref{fig:total_cross_section}, cf. the upper panel. We compare this work to the conventional parameterization of IBD cross sections from Ref.~\cite{Strumia:2003zx}, and prior approaches to radiative corrections from Refs.~\cite{Kurylov:2002vj,Ricciardi:2022pru,Hayes:2016qnu} and from Refs.~\cite{Fayans:1985uej,Vogel:1999zy,Fukugita:2004cq,Raha:2011aa}. QED radiative corrections {\it{shift}} the cross-section predictions {\it{downwards by around $1$-$2\%$}}. The net effect of short-distance electroweak and QCD contributions is approximately of the same size but enters with the opposite sign. In this letter, we update previous compilations of short-distance physics, resulting in percent-level changes in unpolarized cross sections. Outside of the IBD threshold region, deviations to known equations for the long-distance part of QED radiative corrections, after the proper implementation as it is summarized in Ref.~\cite{Raha:2011aa}, are at the level $1$-$2\permil$ mainly due to improvement on radiative corrections in the vicinity of the kinematic end points. We also provide the resulting relative error $\frac{\delta \sigma}{\sigma}$ for the total cross section as a function of the antineutrino energy $E_{\overline{\nu}_e}$ in the lower panel in Fig.~\ref{fig:total_cross_section} and identify the axial-vector-to-vector coupling constant ratio $\lambda$ and the CKM matrix element $V_{ud}$ as dominant sources of the theoretical uncertainty, in agreement with Refs.~\cite{Strumia:2003zx,Ricciardi:2022pru}. The resulting error can be well approximated as $1.85\permil$ across the entire reactor antineutrino energy range. It is dominated by external experimental inputs: $1.71\permil$ and $0.64\permil$ from uncertainties of $g_A$ and $V_{ud}$, respectively.

Our calculation provides a significant improvement in the central value of the total IBD cross section. The results deviate from the commonly used IBD cross-section parameterization from Ref.~\cite{Strumia:2003zx} by around $1$-$2\%$. Such differences are important for understanding the reactor antineutrino anomalies: (i) $\sim6\%$ deficit in overall rates of reactor antineutrinos after averaging over the world data~\cite{Giunti:2021kab} that, however, seems to be resolved by properly accounting for reference spectra of fission electrons~\cite{Estienne:2019ujo,Hayen:2019eop,Li:2019quv,Silaeva:2020msh,Letourneau:2022kfs,Perisse:2023efm,Popov:2023cxd,Zhang:2023zif,Bakhti:2024dcv,Zhang:2025ayc,Giunti:2026uaf}, though it can be still subject to unaccounted uncertainties due to the incompleteness or biases in the nuclear databases; (ii) $\sim10\%$ event excess in the $4$-$6~\mathrm{MeV}$ energy range observed by RENO~\cite{RENO:2015ksa}, Daya Bay~\cite{DayaBay:2015lja,DayaBay:2022eyy,DayaBay:2025ngb}, PROSPECT~\cite{PROSPECT:2022wlf,PROSPECT:2024gps}, STEREO~\cite{STEREO:2020hup,STEREO:2022nzk}, NEOS~\cite{NEOS:2016wee}, Double Chooz~\cite{DoubleChooz:2014kuw,DoubleChooz:2019qbj}, SoLid~\cite{SoLid:2024fgw}, and Neutrino-4~\cite{Serebrov:2020kmd}. Our updated cross sections shift predictions in a direction that could account for $15$-$20\%$ of the observed 5 MeV bump~\cite{Mueller:2011nm,Mention:2011rk,Huber:2011wv,Hayes:2013wra,Hayes:2015yka,Sonzogni:2017wxy,Estienne:2019ujo,Hayen:2019eop,Li:2019quv,Silaeva:2020msh,Letourneau:2022kfs,Perisse:2023efm,Popov:2023cxd,Zhang:2023zif,Bakhti:2024dcv,Zhang:2025ayc}, though further experimental and theoretical inputs are needed for a definitive resolution. Consequently, our refinement of IBD scattering cross sections improves the precise understanding of the reactor antineutrino flux.

\begin{widetext}
\begin{figure*}[htp]
	\centering
	\includegraphics[width=0.85\textwidth]{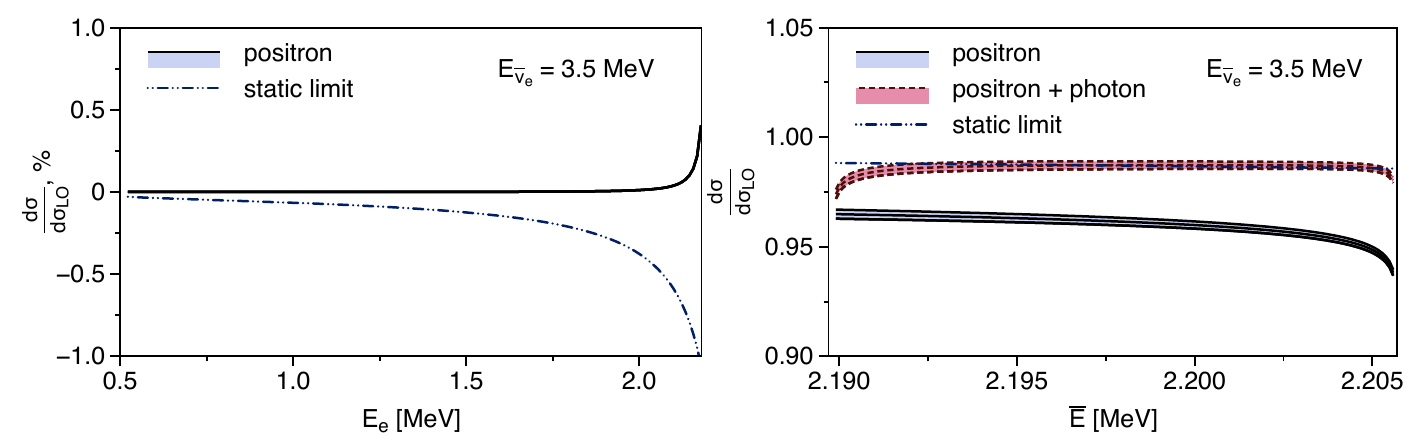}
	\caption{Relative contribution of QED radiative corrections to the positron and electromagnetic energy spectra in IBD is presented for the typical antineutrino energy $E_{\overline{\nu}_e}= 3.5~\mathrm{MeV}$ as a function of the positron or electromagnetic energy. We compare our calculation to the electromagnetic energy spectrum in Refs.~\cite{Fayans:1985uej,Vogel:1999zy,Fukugita:2004cq,Raha:2011aa}, labeled as ``static limit". The left panel illustrates the radiative kinematic region only, while the right panel represents the narrow range of the elastic IBD. $\overline{E}$ refers to either the positron or electromagnetic energy in the right panel. \label{fig:energy_spectra}}
\end{figure*}
\end{widetext}

Next, we compare the resulting single-differential cross sections $\mathrm{d} \sigma$ to the leading-order prediction $\mathrm{d} \sigma_\mathrm{LO}$ for the fixed antineutrino energy as a function of a recoil energy $\overline{E}$, which corresponds to either the positron energy $\overline{E} = E_e$ or the electromagnetic energy $\overline{E} = E_e + E_\gamma$, the sum of the positron $E_e$ and photon $E_\gamma$ energies. We present the ratio $\frac{\mathrm{d} \sigma}{\mathrm{d} \sigma_\mathrm{LO}}$ for such a spectra in Fig.~\ref{fig:energy_spectra} for the representative antineutrino energy at nuclear power plants $E_{\overline{\nu}_e} = 3.5~\mathrm{MeV}$ and compare our results to evaluations of radiative phase-space integrals for the electromagnetic energy spectrum with the no-recoil approximation in Refs.~\cite{Fayans:1985uej,Vogel:1999zy,Fukugita:2004cq,Raha:2011aa}, labeled as ``static limit." In the left panel, we show results for the kinematics allowed only in the radiative process. We notice that previous calculations of radiative corrections to the electromagnetic energy spectrum result in negative radiative cross sections and, therefore, should not be applied in this region of positron energies. In the right panel, we illustrate the spectral distortions in the elastic IBD kinematics. The positron energy spectrum changes significantly within the allowed kinematic range and exhibits logarithmic enhancements below the kinematic limits of the elastic process. By contrast, the electromagnetic energy spectrum remains relatively flat across the elastic kinematics, with logarithmic deviations appearing only near its end points. Besides these logarithmic corrections, the electromagnetic energy spectrum is in perfect agreement with commonly used radiative corrections~\cite{Fayans:1985uej,Vogel:1999zy,Fukugita:2004cq,Raha:2011aa} evaluated in the static limit. By this, we confirm that the analytic result as summarized in Ref.~\cite{Raha:2011aa} is correct and might serve as a first approximation for QED radiative corrections to IBD cross sections in liquid scintillators.

Compared to previous works, we exploit the heavy-nucleon fields both to determine the LECs $g_V$ and $g_A$ and to evaluate QED radiative corrections. It allows us to consistently control contributions to the IBD cross section from the constant terms of the order of $\mathcal{O} \left( \frac{\alpha}{\pi} \right)$. Additionally, we present for the first time the positron energy spectrum, double- and triple-differential distributions, and clarify various types of one-dimensional cross sections. Our results agree with Refs.~\cite{Towner:1998bh,Kurylov:2001av,Kurylov:2002vj,Fukugita:2004cq,Fukugita:2005hs} for both virtual and real QED radiative corrections in the soft-photon approximation, but differ near kinematic end points due to the exact evaluation of radiative phase-space integrals in this work. Moreover, we confirm the collinear region of radiative corrections from Refs.~\cite{Tomalak:2021hec} and~\cite{Tomalak:2022xup}. In agreement with these references, results in the static limit from Refs.~\cite{Vogel:1983hi,Fayans:1985uej,Kurylov:2002vj,Fukugita:2004cq,Raha:2011aa,Ankowski:2016oyj,Tomalak:2021hec,Tomalak:2022xup}, and the Kinoshita-Lee-Nauenberg theorem~\cite{Bloch:1937pw,Nakanishi:1958ur,Yennie:1961ad,Kinoshita:1962ur} in general, our predictions for the electromagnetic energy spectrum are free from collinear singularities. By contrast, the positron spectrum retains a collinear logarithm $\ln(m_e)$ as $m_e \rightarrow 0$, as expected for an observed charged particle~\cite{Dicus:1982bz}.

\section{Conclusion}

We have computed radiative corrections to the inverse beta decay process at reactor antineutrino energies, below $10~\mathrm{MeV}$. For the first time, we consistently included quantum electrodynamics, electroweak, and quantum chromodynamics corrections within the framework of heavy-baryon chiral perturbation theory. To allow an accurate reconstruction of the antineutrino energy with liquid scintillators and achieve subpercent precision goals of JUNO, our calculation improves upon earlier works by employing updated low-energy constants, performing exact phase-space integration for bremsstrahlung, and providing the positron energy spectrum. The results reveal percent-level shifts in the total cross section and characteristic distortions in differential spectra, with theoretical uncertainties below $0.1\%$ level. These predictions are important for current and future reactor neutrino experiments, for supernova neutrino production and detection, and for searches beyond the Standard Model. The methods developed here are applicable to other low-energy charged-current processes.

\section{Acknowledgments}

FeynCalc~\cite{Mertig:1990an,Shtabovenko:2016sxi}, LoopTools~\cite{Hahn:1998yk}, JaxoDraw~\cite{Binosi:2003yf}, Mathematica~\cite{Mathematica}, and DataGraph~\cite{JSSv047s02} were extremely useful in this work. This work is supported by the National Science Foundation of China under Grants No. 12347105 and No. 12447101.

\bibliographystyle{apsrev4-2}

\bibliography{IBD}{}

\end{document}